\documentclass[runningheads]{llncs}
\usepackage[utf8]{inputenc} % allow utf-8 input
\usepackage[T1]{fontenc}    % use 8-bit T1 fonts
\usepackage{hyperref}       % hyperlinks
\usepackage{url}            % simple URL typesetting
\usepackage{booktabs}       % professional-quality tables
\usepackage{amsfonts}       % blackboard math symbols
\usepackage{nicefrac}       % compact symbols for 1/2, etc.
\usepackage{microtype}      % microtypography
\usepackage{xcolor}         % colors
\usepackage{graphicx}
\usepackage{framed}
\usepackage{todonotes}
\usepackage{amsmath}
\usepackage{amssymb}
\usepackage{amsfonts}
\usepackage{algorithm,algpseudocode}
\usepackage{caption}
\usepackage{subcaption}
\usepackage{wrapfig}
\usepackage{todonotes}

\begin{document}
\title{Modeling the Shape of the Brain Connectome via Deep Neural Networks}
\author{Haocheng Dai\inst{1}\thanks{H. Dai and S. Joshi were supported by NSF grant DMS-1912030. P. T. Fletcher was supported by NSF grant IIS-2205417. M. Bauer was supported by NSF grants DMS-1912037, DMS-1953244 and by FWF grant FWF-P 35813-N.}\and
Martin Bauer\inst{2}\and
P. Thomas Fletcher\inst{3}\and
Sarang Joshi\inst{1}}
\institute{University of Utah 
\and Florida State University and University of Vienna
\and University of Virginia }
\authorrunning{H. Dai et al.}
\maketitle              % typeset the header of the contribution
\begin{abstract}
The goal of diffusion-weighted magnetic resonance imaging (DWI) is to infer the structural connectivity of an individual subject's brain in vivo. To statistically study the variability and differences between normal and abnormal brain connectomes, a mathematical model of the neural connections is required.  In this paper, we represent the brain connectome as a Riemannian manifold, which allows us to model neural connections as geodesics. This leads to the challenging problem of estimating a Riemannian metric that is compatible with the DWI data, i.e., a metric such that the geodesic curves represent individual fiber tracts of the connectomics. We reduce this problem to that of solving a highly nonlinear set of partial differential equations (PDEs) and study the applicability of convolutional encoder-decoder neural networks (CEDNNs) for solving this geometrically motivated PDE. Our method achieves excellent performance in the alignment of geodesics with white matter pathways and tackles a long-standing issue in previous geodesic tractography methods: the inability to recover crossing fibers with high fidelity. Code is available at \url{https://github.com/aarentai/Metric-Cnn-3D-IPMI}.
\end{abstract}

\section{Introduction}\label{intro}
Diffusion-weighted magnetic resonance imaging (DWI) enables the non-invasive study of neural connections within the living human brain. 
DWI measures the local diffusion of water within axonal bundles, allowing for local directional estimation of neural connections. Long distance structural connectivity of the brain is then inferred by the process of tractography, which estimates white matter tracts via various streamlining algorithms. Deterministic tractography~\cite{basser2000vivo} computes the integral curves of the vector field associating the most likely direction of fiber tracts with each voxel. However, the simplest deterministic streamline tractography is sensitive to imaging noise and also easily confounded in crossing-fiber regions. Various approaches, such as Kalman filtering~\cite{cheng2014tractography}, probabilistic tractography~\cite{behrens2003characterization}, and front propagation~\cite{sharpening}, have been proposed. The collection of tracts in an individual brain estimated by one or the other methods is referred to as the connectome. 

\noindent{\bf Mathematical models for the shape of the connectome:}
To study the variability in normal populations and to find differences between neural connections in normal and abnormal brains, we need a precise mathematical model of the connectome. Traditionally, individual fiber tracts have been modeled as smooth curves without any intimate link to the underlying geometry of the brain. In geodesic tractography, as proposed by~\cite{inverse,adaptive,adjugate,kaushik2021potential,bihonegn2021geodesic,bihonegn20204th,sharpening}, the brain is modelled as a compact 3D Riemannian manifold, where length-minimizing curves, or {\em geodesics}, represent individual fiber tracts. Recall that a Riemannian manifold is a real, differentiable manifold $M$, equipped with a positive-definite inner product on the tangent space at each point. The shape of the Riemannian manifold (and thus the shape of the geodesics) is determined by the local metric. Smooth Riemannian manifolds with the same topology can have very different shapes because of the differing local metric structure. 

\noindent{\bf Related work on geodesic DWI tractography:}
DWI is the foundation to model an individual brain as a Riemannian manifold. With the Riemannian-metric-equipped manifold, we can infer the white matter pathways and also the shape of an individual's connectome. O'Donnell et al.~\cite{inverse} first proposed the geodesic tractography algorithm that uses the inverse of the diffusion tensor as the Riemannian metric and treats geodesic curves under the metric as white matter pathways. However, there is a tendency in the inverted-tensor metric for geodesics to easily deviate from the principal eigenvector directions in high-curvature areas. To address this issue, Fletcher et al.~\cite{sharpening} enhanced the metric by ``sharpening'' the inverted-tensor metric, i.e., taking the eigenvalues of the metric tensor to some power so as to increase the anisotropy. But this strategy does not take into account the spatially varying curvature of the vector field, and it increases the sensitivity to noise. Fuster et al.~\cite{adjugate} demonstrated that using the adjugate of the diffusion tensor field as the Riemannian metric gave improved geodesic tractography over the inverted sharpened metric while being more robust to imaging noise. In order to strengthen the adherence of geodesics to the white matter pathways, Hao et al.~\cite{adaptive} developed an adaptive Riemannian metric by applying a conformal scalar field to the inverse of the diffusion tensor, which necessitates solving a Poisson equation on the Riemannian manifold. Campbell et al.~\cite{campbell2021} further advanced the Riemannian formulation of structural connectomes by introducing methods for diffeomorphic image registration and atlas building using the Ebin metric on the space of Riemannian metrics. The significant advantage of the Riemannian geometric framework is that it enables the formulation of atlas building as a statistical Fr\'{e}chet mean estimation problem. Furthermore, the entire toolbox of geometrical statistics can now be applied to the statistical analysis of populations of connectomes, which addresses a current challenge in neuroscience --- how to statistically quantify the variability of human brain connectivity and differences in the connectome across populations. 

\noindent{\bf Contributions:}
The principal aim of this work is to innovate the main building block of the Riemannian formulation for structural connectome atlas building: the estimation of a Riemannian metric, such that the corresponding geodesic tractography provides a faithful description of the tractogram, i.e., that the geodesics of the Riemannian metric follow the integral curves of a set of given vector fields (representing fiber orientations). The existing Riemannian metric estimation techniques for DWI data exhibit two major limitations: the accuracy of the Riemannian metric estimation, i.e., alignment of the geodesics with the fiber tracts, and the fact that they are all based on a single DTI model and are thus not able to consider multiple vector fields simultaneously, which limit their ability to model crossing fibers appropriately. More modern modeling techniques, such as HARDI~\cite{tuch2002high}, Q-Ball~\cite{tuch2004q} and DSI~\cite{tuch2002diffusion}, are able to infer multiple fiber directions at each point in the brain. 

In this paper, we show for the first time how one can leverage deep neural networks (DNNs) to estimate a metric structure of the brain that can accommodate fiber crossings, i.e., multiple fiber directions, which is a natural modeling tool to infer the shape of the brain from DWI. We reduce the problem of estimating a Riemannian metric given tractography estimated from DWI, to that of solving a highly nonlinear set of partial differential equations (PDEs) of the form $\mathcal{L}g(x)=f(x), x\in \Omega \subset \mathbb R^{n}$, where $f(x)$ is the given data, and  $\mathcal{L}$ is a non-linear differential operator. This allows us to leverage deep learning frameworks that have been proposed for solving such PDEs precisely, where we use convolution encoder-decoder neural networks (CEDNNs)~\cite{denseed} for representing the solution space. CEDNNs use the universal approximation property of fully-connected convolutional networks to approximate the solution function $g(x)$. The weights of the network are estimated to minimize a cost function which incorporates the PDE in a self-supervised manner and is usually of the form $\|\mathcal{ L}g(x)-f(x)\|$. CEDNNs are particularly suited when the input data domain is a structured grid such as in the DWI application. 

Using this network architecture and spatially discretized vector fields from any of the plethora of models for local fiber directions, our method achieves excellent performance in terms of geodesic-white-matter-pathway alignment. In particular, we show that the proposed method outperforms any of the previously proposed methods in Riemannian metric estimation for geodesic tractography. In addition to simple deployment and boundary insensitivity, our approach also tackles the long-standing issue in previous methods: the inability to recover crossing fibers with high fidelity. Towards this aim, we exhibit that the metric estimation is able to faithfully represent multiple vector fields as geodesic vector fields of the estimated metric. We inherit the validity of the tracts from the choice of the preferred local directional estimation algorithm, which is explicitly not the focus of this work. The algorithms presented herein are a part of the overall program to study the human brain and its variability in populations. 

\section{Estimating Riemannian Metrics from Geodesics}
In this section, we will introduce a new inverse problem that will
be at the center of our approach: the estimation of a Riemannian metric based on the observation of (possibly) multiple fiber directions as the tangents to geodesic curves. We will first recall some definitions and concepts from Riemannian geometry. For further details, we refer to classic textbooks such as~\cite{do1992riemannian}. In all of this work, our modeling space is a finite-dimensional manifold $M$ (possibly with boundary). In our application, the topology of the manifold $M$ will be rather trivial and thus we assume in the following that $M$ is a bounded subset of $\mathbb R^n$ with $n\in \{2,3\}$. 

Next we introduce the concept of an integral curve: given a vector field $\mathbf{v}\in \mathfrak X(M)$, i.e., a map from $M$ to $TM$, we call a curve $\gamma: \mathbb R \to M$ an integral curve of $\mathbf{v}$ if $\partial_t\gamma(t) = \mathbf{v}(\gamma(t))$, i.e., the curve follows the flow lines of the vector field. A Riemannian metric $g$ on $M$ is a family of inner products on each tangent space $T_xM$ that depend smoothly on the base point, $x \in M$. Note that in local coordinates we can identify the Riemannian metric with a field of positive-definite, symmetric matrices $g(x)$, and the inner product between two tangent vectors $v,w\in T_xM$ is simply given by $\langle v,w\rangle_{g} = v^T g(x)w$. 
We call a curve between $p$ and $q$ a (minimizing) geodesic if it minimizes the length functional 
$
    L(\gamma)=\frac12\int^1_0 \sqrt{\langle \partial_t \gamma,\partial_t \gamma\rangle_{g_\gamma}}dt,
$
where $g_\gamma$ denotes the Riemannian metric at $\gamma(t)$. %Locally length-minimizing curves satisfy an ordinary differential equation (ODE), called the geodesic equation, which is the first-order optimality condition $dL(\gamma)=0$. 
For every Riemannian metric there exists a unique connection $\nabla^g$, called the Levi-Civita covariant derivative, which encodes this notion of geodesic curves, i.e., a curve $\gamma$ is a geodesic if and only if it satisfies the equation $\nabla^g_{\mathbf{v}} \mathbf{v}=\sigma\mathbf{v}$, where $\mathbf{v} = \partial_t\gamma,$ and $\sigma=\langle\mathbf{v},\nabla^g_\mathbf{v} \mathbf{v}\rangle_{g}/||\mathbf{v}||^{2}_{g}$. We call a vector field a unit geodesic vector field if all its integral curves are geodesics with constant speed, i.e., $\nabla^g_\mathbf{v} \mathbf{v}=0$. Integral curves of both geodesic as well as unit geodesic vector fields are length minimizing --- they only differ in their parameterization along the curve. Given a vector field $\mathbf v$ we aim to find a Riemannian metric $g$ such that $\mathbf v$ is a geodesic vector field. The question under what conditions such a Riemannian metric exists has been intensively investigated, see e.g.~\cite{rechtman2010existence}. Note that if $\mathbf{v}$ is a non-vanishing geodesic vector field of a metric $g$, then $\mathbf{v}/\|\mathbf{v}\|_{g}$ is a unit geodesic vector field. We found, however, that our model is significantly harder to optimize, when insisting on unit geodesic vector fields.

We are now able to formulate the inverse problem studied in this paper as:
\begin{framed}
\noindent {\bf Regularized, inexact metric estimation:} Given vector fields $\mathbf{v}_i\in \mathfrak X(M)$, $i\in \{1,\ldots,m\}$, find the Riemannian metric $g$ on $M$ that minimizes the energy functional
\begin{equation}\label{loss}
\mathcal E(g)=\sum_{i=1}^{m} \|\nabla^g_{\mathbf{v}_i} \mathbf{v}_i-\sigma_i\mathbf{v}_i\|_{2}+\alpha \operatorname{Reg}(g), 
\end{equation}
where $\alpha>0$ is a weight parameter. 
\end{framed}
Here the first term enforces the condition that the vector fields $\mathbf{v}_i$ are (close to being) geodesic vector fields, while the second term is a regularization parameter that is responsible for the solution selection.  %We note that this formulation relates directly to the PINN\todo{Maybe remove PINNs?} formulation for solving PDEs using neural networks. 
We have investigated several regularization terms, such as the Frobenius norm of the difference to the Euclidean metric. In our experience, adding this explicit regularization terms did not improve the performance of the algorithm, which suggests that the implicitly regularization properties via the solution parameterization as a neural network are sufficient for the proposed application. 

To get a better understanding of the above loss function, we can write a coordinate expression of $\nabla^g$ for a vector field $\mathbf{v}$ as:
\begin{equation}\label{nabla}
    \nabla^g_{\mathbf{v}}\mathbf{v}=\sum_k\left(\sum_i v^i\frac{\partial v^k}{\partial x^i}+\sum_{i,j}\Gamma^k_{ij}v^iv^j\right)\mathbf{e}_k,
\end{equation}
where $\mathbf{v}=v^i\mathbf{e}_i$ with $\mathbf{e}_i=\frac{\partial}{\partial x^i}$ being the $i$-th basis vector. Furthermore, $\Gamma^k_{ij}$ are the Christoffel symbols, which are defined as
$
    \Gamma^k_{ij}=\frac{1}{2}\sum^n_{l=1}g^{kl}\left(\frac{\partial g_{jl}}{\partial x^i}+\frac{\partial g_{il}}{\partial x^j}-\frac{\partial g_{ij}}{\partial x^l}\right),
$
where $g_{ij}$ denotes the entries of the Riemannian metric $g$, and $g^{ij}$ represents the entries of the inverse of metric $g^{-1}$. 

\section{Algorithms and Implementation}

\noindent{\bf Metric Estimation via Neural Networks:} We will now present a novel deep learning framework~\cite{denseed} for solving the inverse problem formulated above, which employs a convolutional encoder-decoder neural network (CEDNN) approach to construct the multi-scale features from high-dimensional input. By wrapping the input vector fields and the output Riemannian metric into the loss function, the network is trained to capture the heterogeneous mapping between the given  vector fields and the resulting solution (metric). CEDNN takes the whole spatially discretized vector fields as the input and outputs the entire metric field which minimizes Eq.~\eqref{loss}. 

The CEDNN architecture  adopts the dense block~\cite{densenet} paradigm, which furthers the ideas of residual learning in ResNet~\cite{resnet} and bypassing paths in highway networks~\cite{srivastava2015training} by concatenating every previous layer's output as the input of the current layer in a feed-forward fashion. The dense connectivity in the dense block improves the information flow in the network, without introducing any optimization difficulty. The encoder contracts the higher-level context and feature of the input, while the decoder commits to recovering the location information to the same scale as the original input fields. 

\begin{figure}
\centering
% \vspace{-20pt}
\includegraphics[width=1\textwidth]{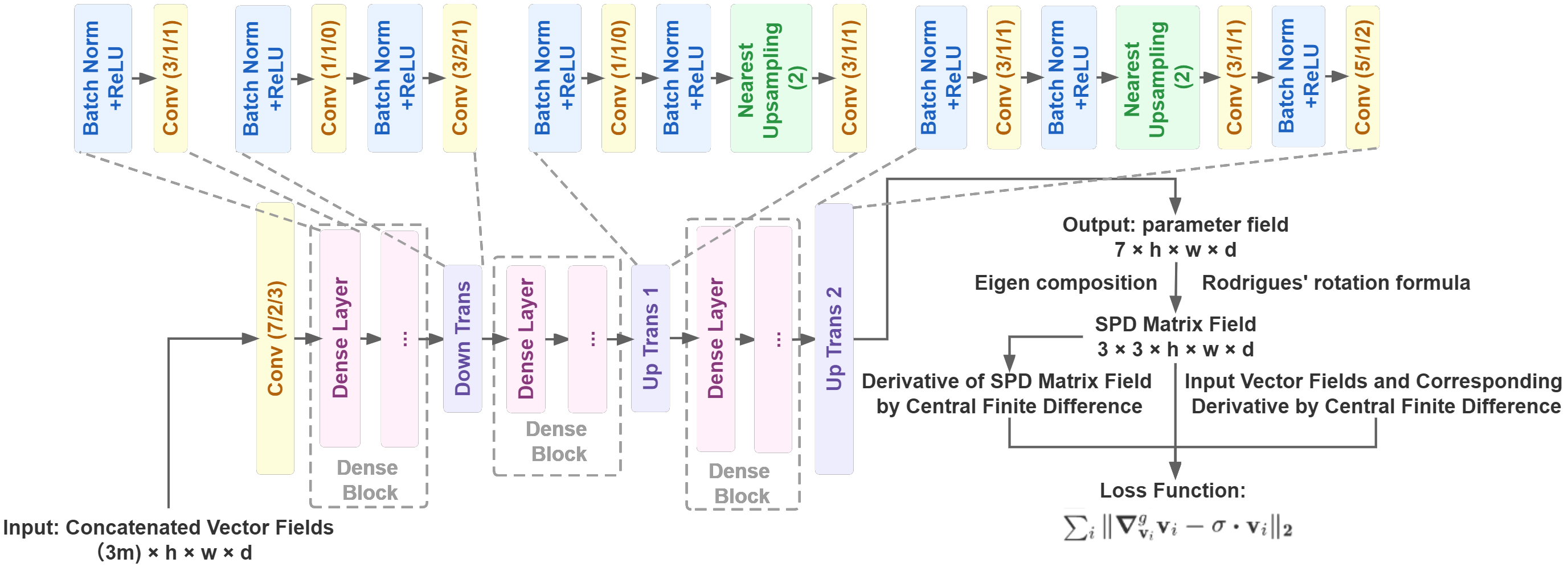}
\caption{The architecture of the proposed convolutional encoder-decoder neural networks for 3D solution. Here, $h,w,d$ denote the shape of the input vector fields, and $m$ represents the number of total vector fields. The numbers in a \textit{Conv} box stand for the kernel size, stride, and padding of the convolution, respectively. The number in a \textit{Nearest Upsampling} box indicates the scaling factor.} \label{arch}
% \vspace{-20pt}
\end{figure}

Our CEDNN implementation takes an $m\times n$-channel input, where $n$ is the dimension of the vector fields and $m$ is the number of distinct geodesic vector fields $\mathbf{v}_i$ used for the metric estimation. By sending the concatenated vector fields into the network, CEDNN yields an output of $((n+1)\times n/2+1)$-channel tensor. We form the final estimated metric, the $n\times n$ symmetric positive-definite (SPD) matrix $g$, through eigencomposition: $g=\mathbf {R}\mathbf {\Lambda}\mathbf {R}^T$, where the rotation matrix $\mathbf {R}$ follows Rodrigues' rotation formula: $\mathbf {R}=\mathbf {I} +(\sin \theta )\mathbf{K} +(1-\cos \theta )\mathbf {K} ^{2}$, $\mathbf {\Lambda}$ is a diagonal matrix with positive diagonal entries, $\mathbf{I}$ is an identity matrix, and $\mathbf{K}$ is a skew-symmetric matrix. $\mathbf{R},\mathbf{\Lambda},\mathbf{K},\mathbf{I}\in\mathbb{R}^{n\times n}, \theta\in\mathbb{R}$ and we enforce the diagonal entries in $\mathbf{\Lambda}$ to be positive through exponential function. $\mathbf{\Lambda},\mathbf{K},\theta$ are respectively parameterized by $n$, $n\times(n-1)/2$ and 1 real numbers, the sum of which is equivalent to the channel number of the output tensor. This formulation expedites the training speed by 5 folds, compared to forming a SPD matrix via built-in \texttt{torch.matrix\_exp}. To compute spatial gradients, we adopt a central finite-difference scheme to approximate the derivatives in the loss function as given in Eq.~\eqref{loss}. For more details on the architecture and workflow, see Fig.~\ref{arch}.

\noindent{\bf Geodesic Tractography:}
Once we have estimated a neural network representation of the metric $g$, we are able to calculate the Christoffel symbols and then integrate the geodesic equation (Eq.~\eqref{nabla}), which directly leads to the corresponding geodesic tractography. Recall that in three-dimension situation, the geodesic equation~\eqref{nabla} can be written as the following system of second-order ODEs:
$
\partial_t^2 \gamma^{(k)}(t)+\sum_{i,j=1}^{3} \partial_t\gamma^{(i)}(t)\cdot\Gamma^k_{i,j}(\gamma(t)) \cdot\partial_t\gamma^{(j)}(t)=\sigma\mathbf{v}^{(k)}(\gamma(t)),
$
where $\partial_t^2 \gamma^{(k)},\partial_t \gamma^{(k)}$ are the $k$-th components of the acceleration and velocity vectors and where  $\Gamma^k_{ij}(\gamma(t))$ are the Christoffel symbols evaluated at position $\gamma(t)$. Geodesic shooting solves this second-order ODE with a set of initial conditions: the starting position of the geodesic $\gamma(0)$ and its starting velocity $\partial_t \gamma(0)$. %We use a fourth-order Runge-Kutta scheme for integrating the above second-order ODE. 

\section{Experiments}

In all of the experiments presented below, the Adadelta optimizer was employed to minimize the loss function. We also experimented with various other optimizers, including AdaGrad, Adadelta, and Adam. In our experience, the Adadelta optimizer was consistently superior.
All computations were carried out on an Nvidia Titan RTX GPU. Under this setting, a 2D example took less than 3 minutes (3,000 iterations) to achieve convergence, and a 3D example took less than 30 minutes (10,000 iterations) for convergence. 

\begin{figure}[h]
  % \vspace{-20pt}
     \centering
     \includegraphics[width=0.26\textwidth]{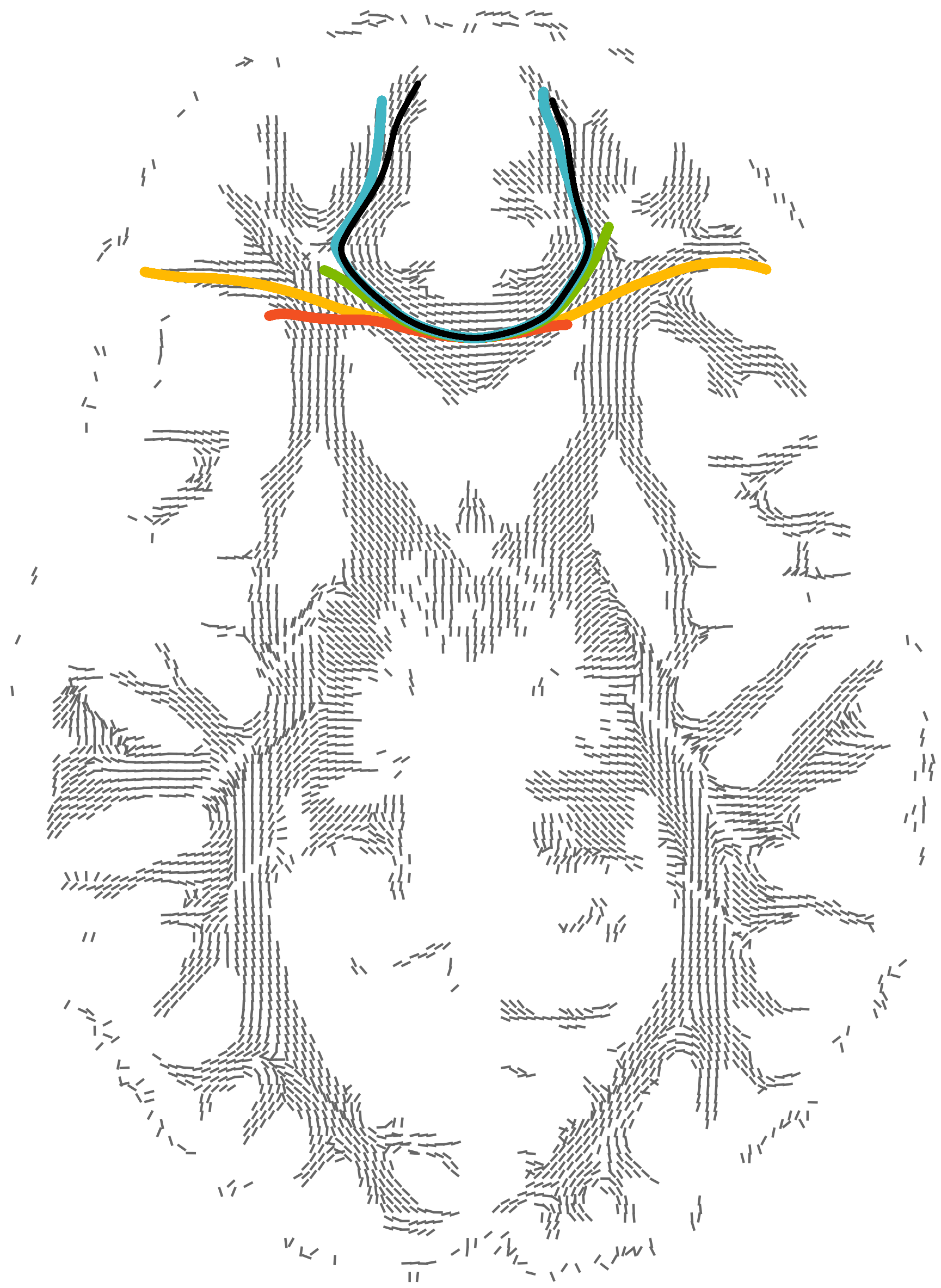}
     \includegraphics[width=0.35\textwidth]{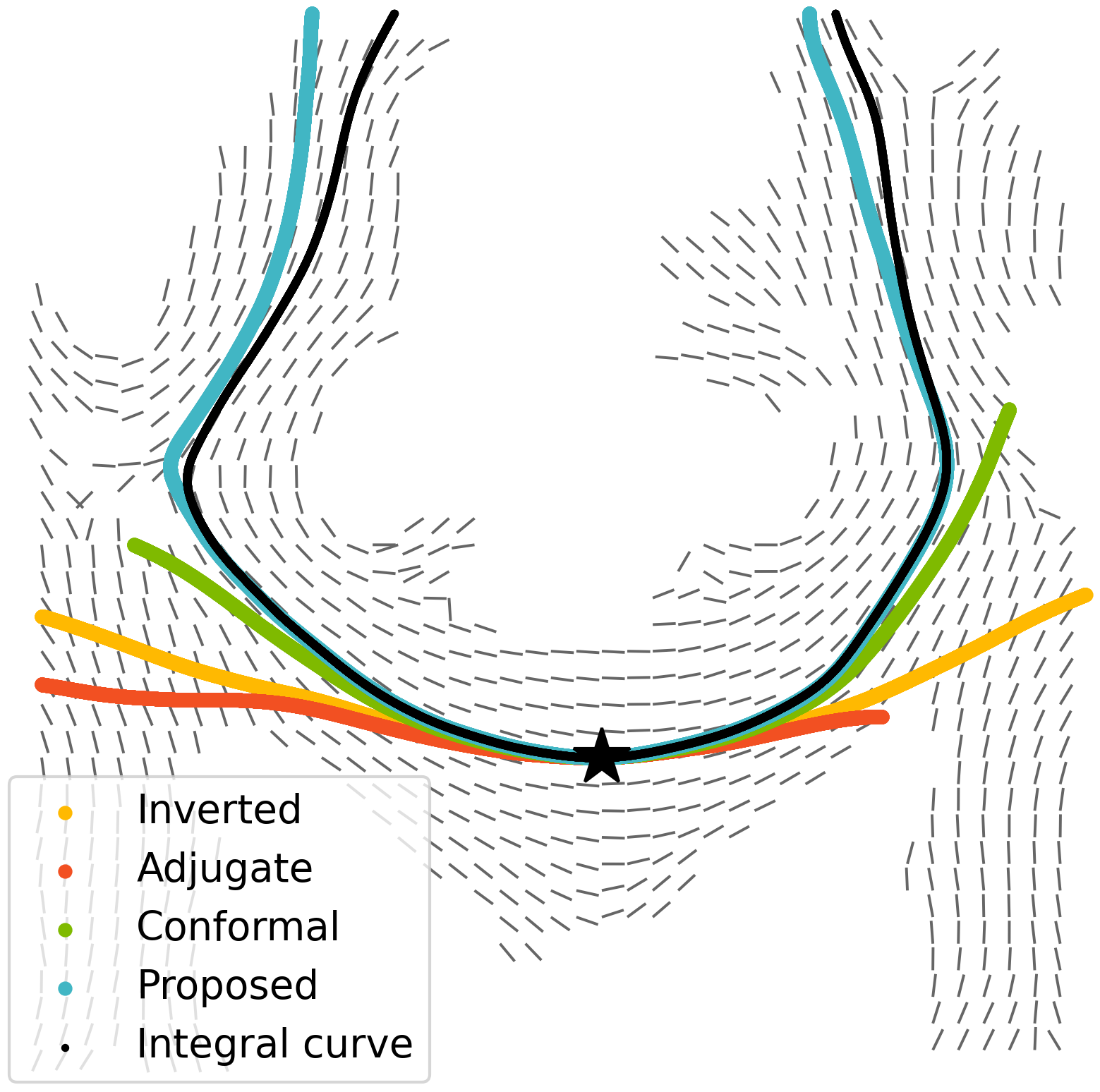}
     \includegraphics[width=0.36\textwidth]{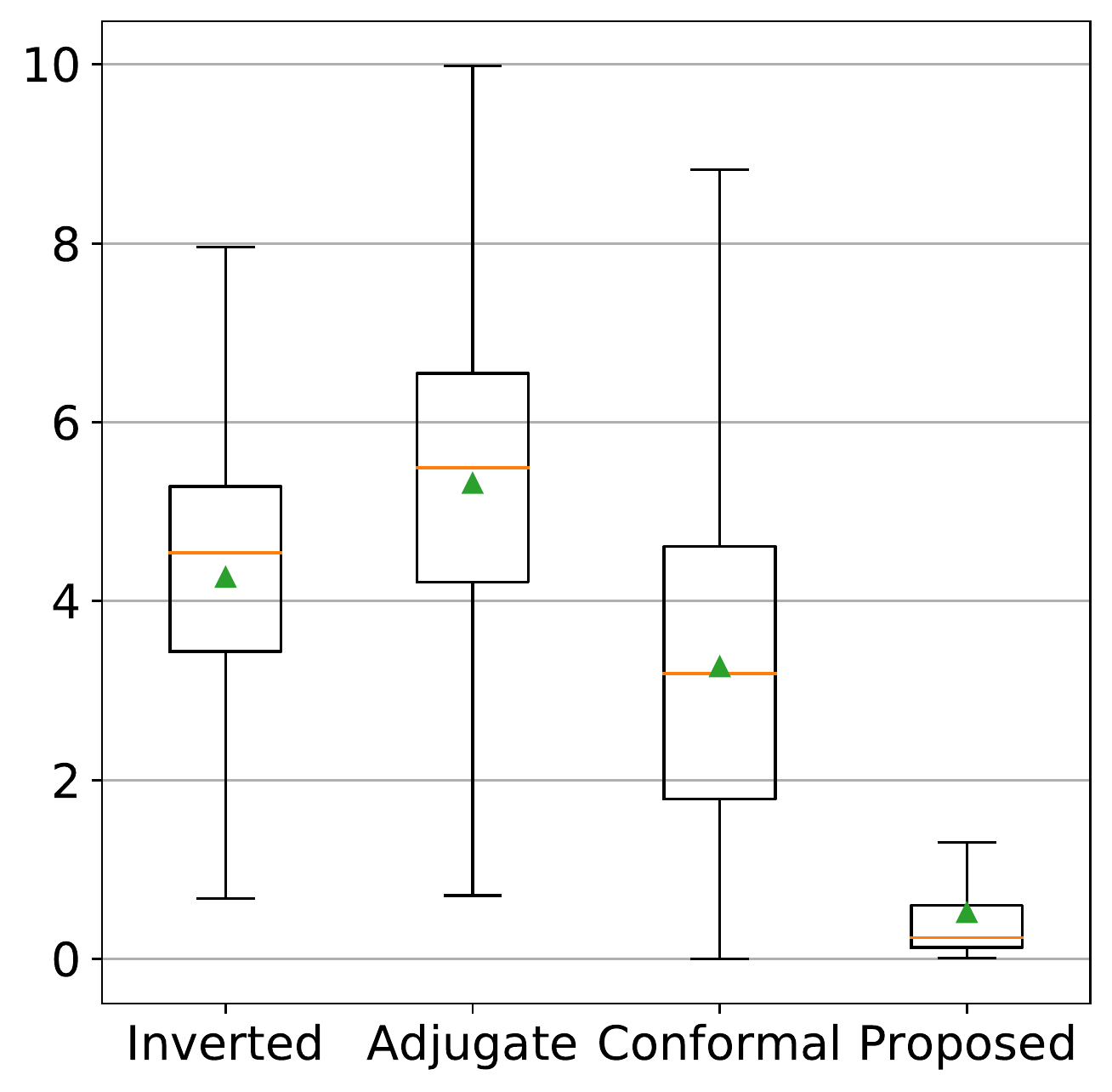}
        \caption{Left: axial view of a projected vector field derived from HCP 100610 DWI. Central: detailed view of left panel, the geodesics and integral curve start from the star. Right: Box plot of mean min errors between the integral curve and geodesics generated by different methods over 38 HCP brain subjects. Green triangles stand for the mean of the mean min errors, and orange bars represent the median of the mean min errors.}
        \label{fig:compare}
  % \vspace{-20pt}
\end{figure}
\subsection{Validation and Comparison: 2D Brain Slices} 
In our first experiment, we compared the geodesic-white-matter-pathway alignment of the proposed method to state-of-the-art geodesic tractography methods: the \textbf{inverted} diffusion tensor metric~\cite{inverse}, the \textbf{adjugate} of the diffusion tensor~\cite{adjugate}, and the \textbf{conformal} metric~\cite{adaptive}. 
We want to emphasize that all these baseline methods are approximating the metric based on a single diffusion tensor image, which yields only one corresponding vector field. Consequently, the crossing fiber estimating ability will not be tested in this section. For easy visualization and interpretability, we performed the comparison on projected 2D brain slices from the Human Connectome Project (HCP)~\cite{hcp}, cf. Fig.~\ref{fig:compare}. The left and central panels in Fig.~\ref{fig:compare} demonstrate an example of geodesics shooting from a seed point in the genu of the corpus callosum. The geodesics derived by the other three methods deviate from the integral curve eventually, while ours provides a significantly better alignment to the ground truth. 

% \begin{wraptable}{r}{5cm}
%   \vspace{-15pt}
%   \caption{Mean and median of mean min errors across different methods over 38 HCP subjects. }
%   \label{table:compare}
%   \centering
%   \begin{tabular}{ccc}
%     \toprule
%     Methods   & Mean & Median \\
%     \midrule
%     Inverted  & 4.2660 & 4.5393\\
%     Adjugate  & 5.3153 & 5.4906\\
%     Conformal & 3.2658 & 3.1915\\
%     Proposed  & \textbf{0.5172} & \textbf{0.2388}\\
%     \bottomrule
%   \end{tabular}
%   \vspace{-15pt}
% \end{wraptable}

To quantitatively measure the geodesic-white-matter-pathway alignment over these methods, we tested these algorithms on brain slices from 38 HCP brain subjects. On each brain slice, we uniformly cast 400 seed points in the genu of corpus callosum region, where all the seed points are chosen to be non-grid points with the corresponding vectors being obtained by bi-linear interpolation, i.e., these vectors have never been seen by the network. We then integrated the geodesic and integral curve from each seed point and calculate the error of the geodesic to the corresponding integral curve. To calculate the error of curve $Q$ to curve $P$, we view $P,Q$ as finite point sets and consider the mean min error between these two sets as
$
    \operatorname{Error}(P,Q)=\frac{1}{|P|}\sum_{p\in P}\min_{q\in Q} \|p-q\|_2,
$
where $P$ denotes an integral curve and  $Q$ a geodesic tractography curve starting at the same seed point. The boxplot in Fig.~\ref{fig:compare} visualizes the distribution of 15,200 integral-curve-geodesic error sample points across 38 HCP subjects. The boxplot demonstrates our method outperforms the other three methods in terms of both mean and median of mean min errors by a large margin. 

\subsection{2D Synthetic ``Braid''}\label{sec:comparison}
In this section, we aim to compare the metric estimation ability of CEDNNs with a baseline physics-informed neural networks (PINNs)~\cite{pinn} implementation, where we pay particular attention to the ability of representing crossing fibers. The baseline PINNs use multilayer perceptrons (MLPs) to represent the solution space.
%Irrespective of sending the whole vector field as the input, we feed spatial coordinates $x\in \mathbb{R}^n$ as batch into the network and expect it yields a batch of metric tensor $g\in \mathbb{R}^{n\times n}$ at the corresponding location. 
The loss function in the baseline PINNs is formulated the same way as in the CEDNN implementation, where the  derivative of metric w.r.t. the spatial coordinates was computed by taking advantage of the automatic differentiation engine \texttt{autograd} in \texttt{PyTorch}.

\begin{figure}
%   \vspace{-20pt}
     \centering
     \begin{subfigure}{0.23\textwidth}\label{cos}
         \centering
         \includegraphics[width=\textwidth]{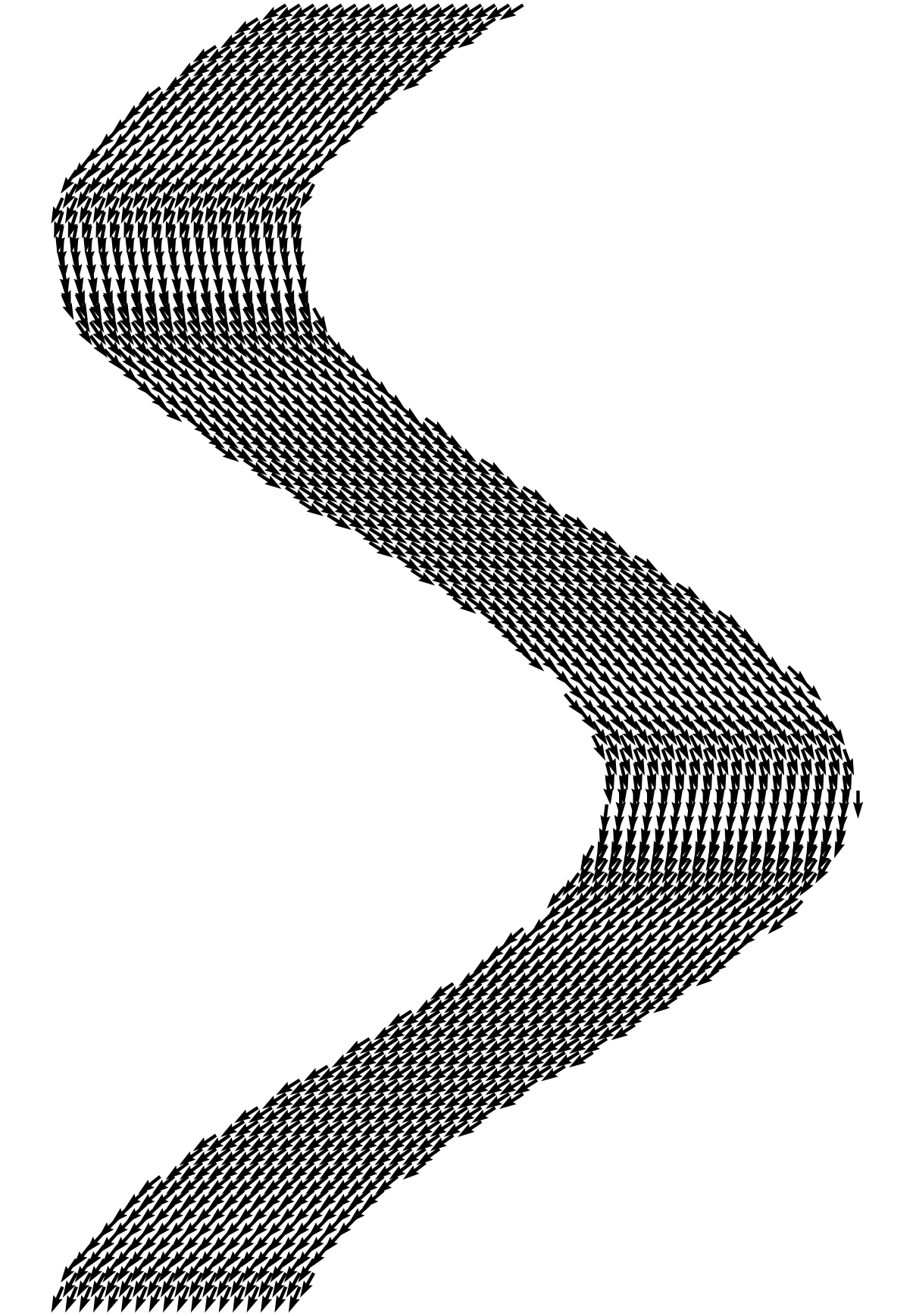}
         \caption{vector field $\mathbf{v}_1$}
     \end{subfigure}
     \begin{subfigure}{0.23\textwidth}\label{sin}
         \centering
         \includegraphics[width=\textwidth]{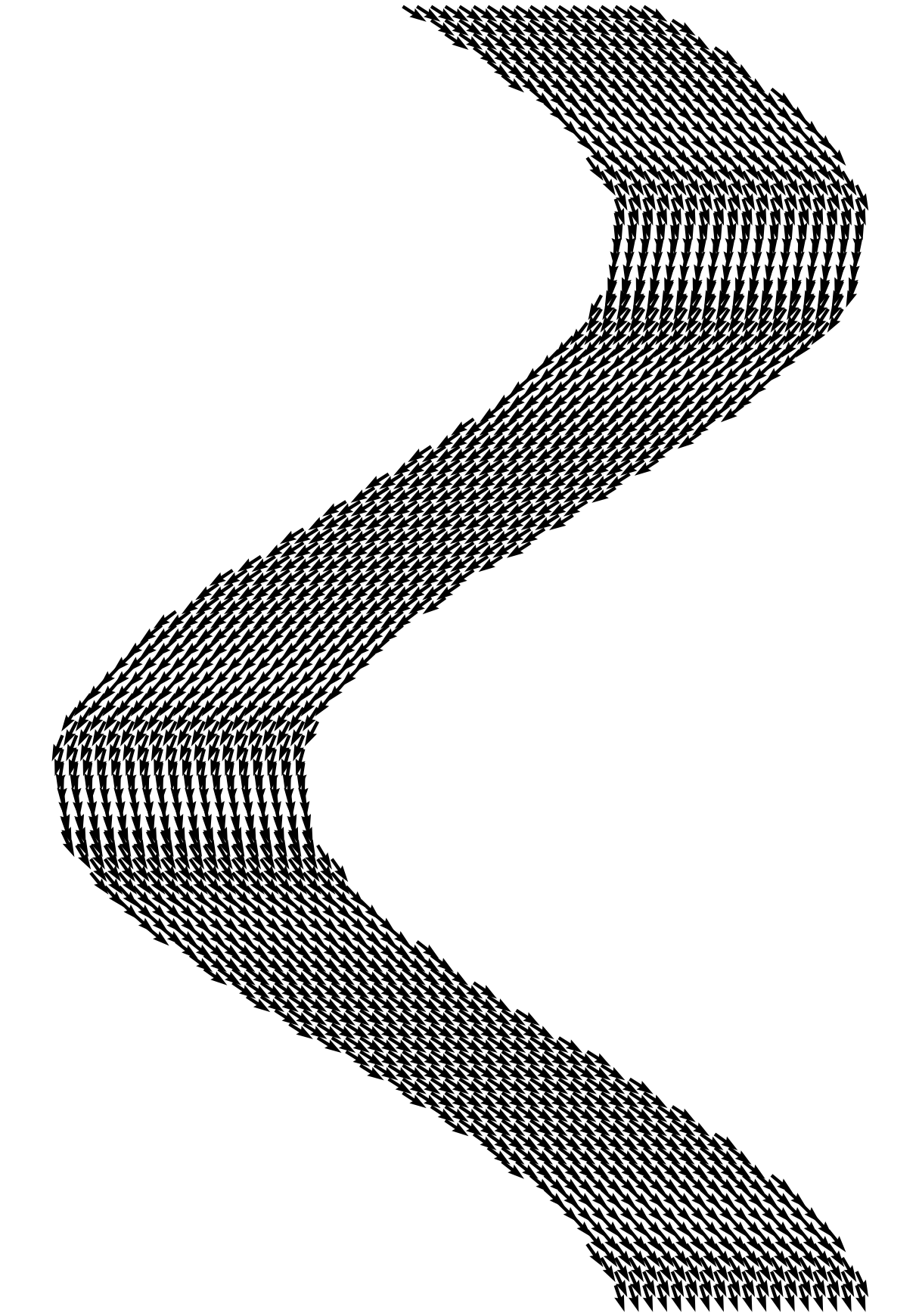}
         \caption{vector field $\mathbf{v}_2$}
     \end{subfigure}
     \begin{subfigure}{0.22\textwidth}
         \centering
         \includegraphics[width=\textwidth]{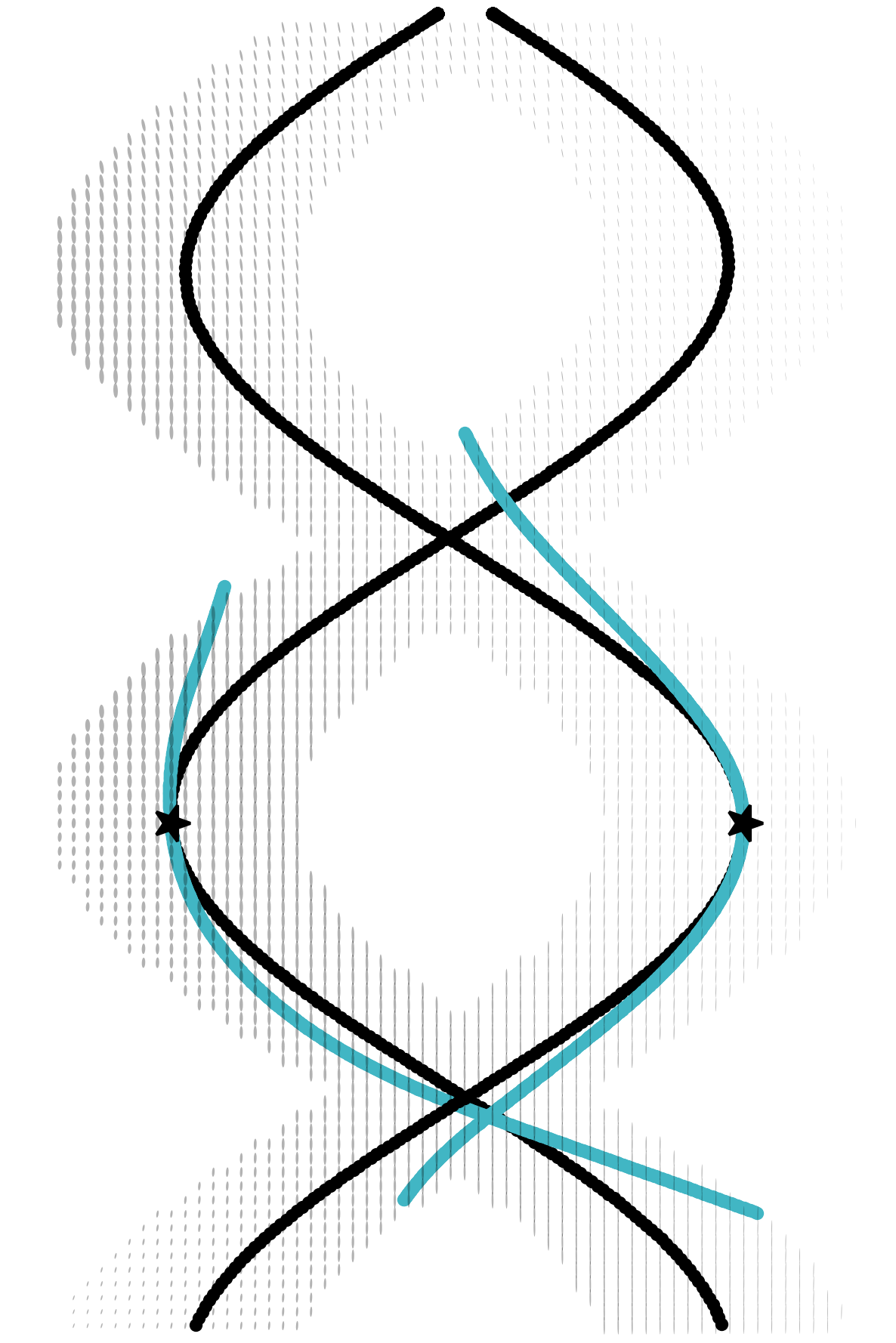}
         \caption{metric $g^P$}
     \end{subfigure}
     \begin{subfigure}{0.22\textwidth}
         \centering
         \includegraphics[width=\textwidth]{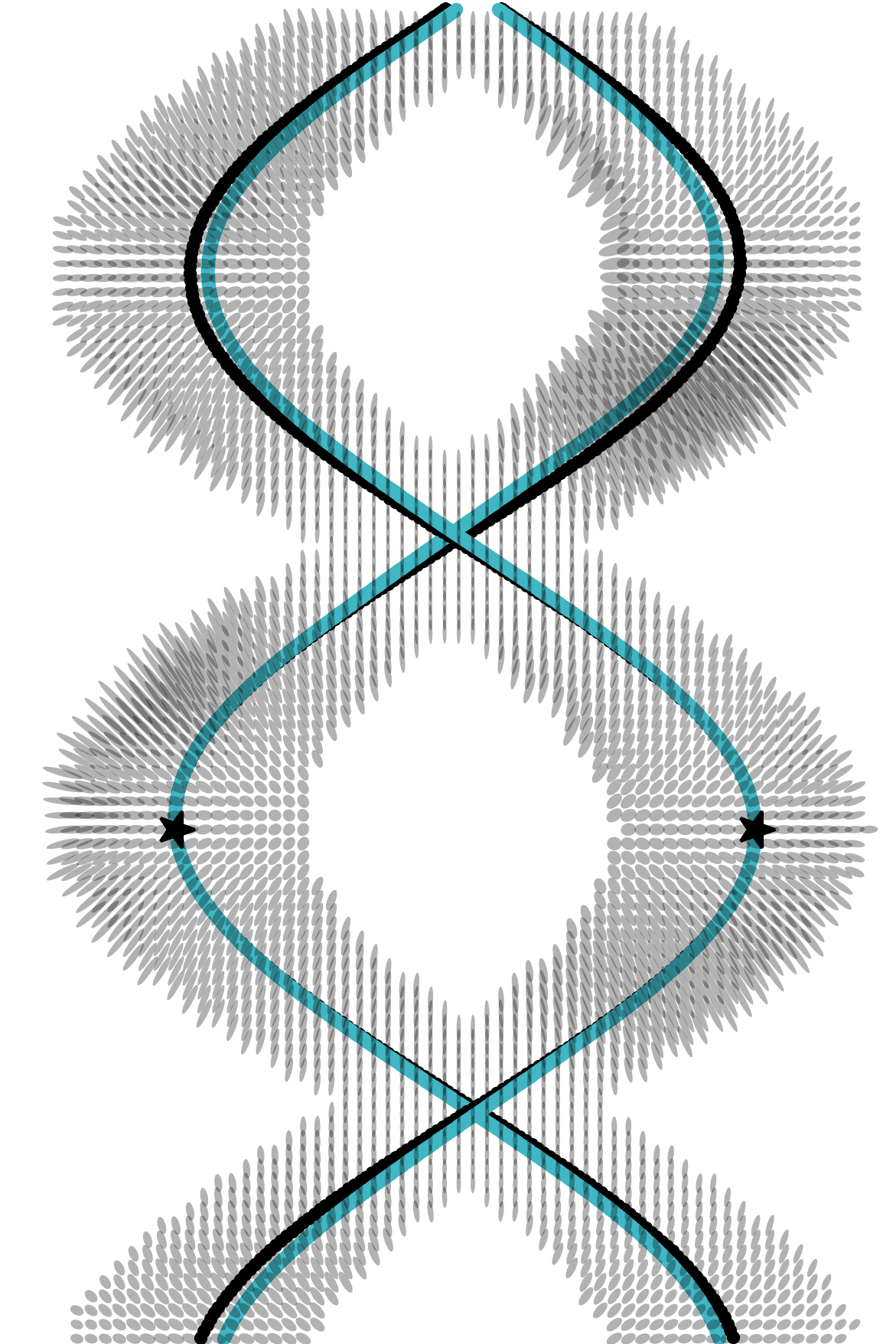}
         \caption{metric $g^C$}
     \end{subfigure}
    \caption{(a) input vector field $\mathbf{v}_1$; (b) input vector field $\mathbf{v}_2$; (c) integral curves (black) running on input vector fields and geodesics (indigo) running on PINN-estimated Riemannian metric field $g^P$ at iteration 1000 (background ellipses represent metric tensors), shooting from seed points (star) with the same initial velocity vector as the corresponding integral curve; (d) integral curves (black) and geodesics (indigo) on CEDNN-estimated Riemannian metric field $g^C$ at iteration 1000 (background ellipses represent metric tensors), shooting from seed points (star) with the same initial velocity vector as the corresponding integral curve.}
    \label{fig:braid}
%   \vspace{-20pt}
\end{figure}

For the experimental data, we synthesized two vector fields in a ``braid'' pattern of two intertwining pathways (see (a) and (b) in Fig.~\ref{fig:braid}). The central integral curves of the vector bundle are two trigonometric functions: $x_2=20\cos(\frac{1}{4\pi}(x_1-60))+50$ and $x_2=20\sin(\frac{1}{4\pi}x_1)+50$, where $x_1,x_2$ are spatial coordinates. We then constructed the curve bundle by translating the central integral curve across nine pixels horizontally. The vector field is generated by calculating the tangent vector of the curves at each point, making the curves integral to the vector field. The aim is to estimate a Riemannian metric field such that these curves are geodesic curves.

\begin{wrapfigure}[19]{r}{0.5\textwidth}
  \vspace{-20pt}
  \begin{center}
     \centering
     \includegraphics[width=0.45\textwidth]{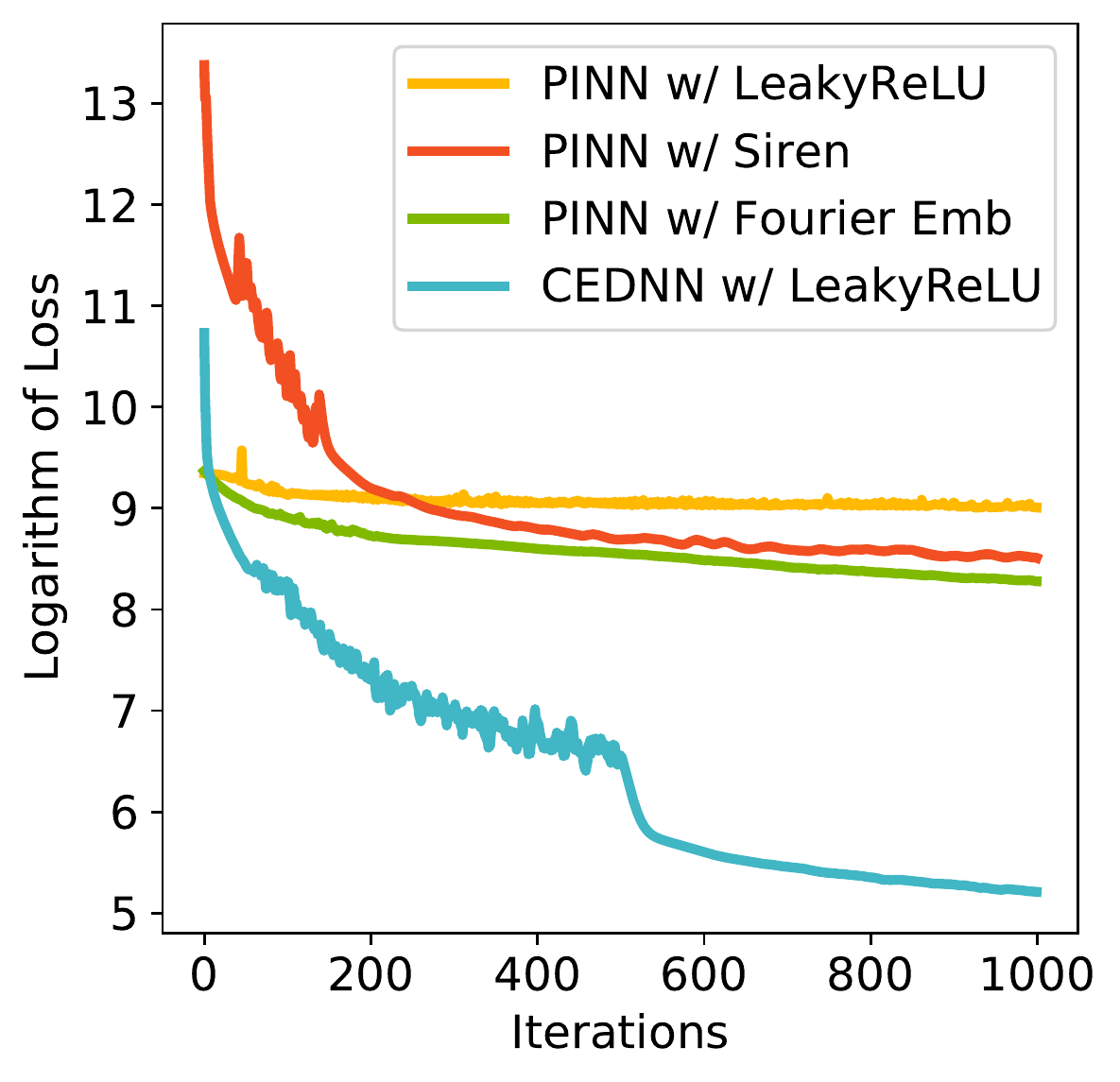}
  \end{center}
  \vspace{-15pt}
  \caption{ Log-scaled (base 10) loss convergence comparison between PINN with LeakyReLU activation, with Siren activation, with Fourier embedding and CEDNN with leakyReLU on 2D synthetic ``Braid''.}
\label{fig:loss_trend}
% \vspace{-20pt}
\end{wrapfigure}

The CEDNN in our experiment is configured with the following hyper-parameters: the number of dense layers in the three dense blocks are 6, 8, 6, with a growth rate of 16, thereby leading to a total of 747,147 parameters. We used an initial learning rate of $1\times 10^{-4}$ for the optimization. 
Fig.~\ref{fig:loss_trend} presents the loss $\sum_{i=1}^{m} \|\nabla^g_{\mathbf{v}_i} \mathbf{v}_i-\sigma_i\mathbf{v}_i\|_{2}$ on a log-scale at each iteration. The baseline PINN uses an approximately equivalent amount of parameters as the CEDNN. PINNs with different sizes were extensively explored, however, we did not observe any salient difference in final loss brought by these configurations. In addition, we configured the PINN with Fourier embedding~\cite{tancik2020fourier} and Siren activation functions~\cite{sitzmann2019siren}, which boosted the performance of the PINN in a considerable magnitude, yet still underperforming CEDNN by orders of magnitudes: in Fig.~\ref{fig:loss_trend}, it is noticeable that the CEDNN converges much faster to a significantly lower residual loss despite the fact that PINN enjoys about the same amount of parameters as the CEDNN. This enforces the conclusions of ~\cite{krishnapriyan2021characterizing} that the limitation exhibited by PINNs is due to optimization difficulties --- irrespective of expressibility of the solution. See also \cite{chuang2022experience}, which reports several pitfalls in using PINNs for fluid dynamic simulations. 

Fig.~\ref{fig:braid} shows the metric fields $g^P$ (Fig.~\ref{fig:braid}(c)) and $g^C$ (Fig.~\ref{fig:braid}(d)) estimated by PINN and CEDNN respectively and the alignment of ground truth integral curves (black) and geodesics (indigo) associated with the estimated metric: the geodesics on the metric $g^C$ align notably closer to the ground truth integral curves than the one on $g^P$. In addition to the excellent integral-curve-geodesic alignment, the CEDNN-estimated metric behaves as expected even at the crossing region --- the geodesics are not confounded at the crossing. 

\subsection{3D Brains and Crossing Fibers}
\begin{figure}[htbp]
%   \vspace{-15pt}
     \centering
     \includegraphics[width=0.21\textwidth]{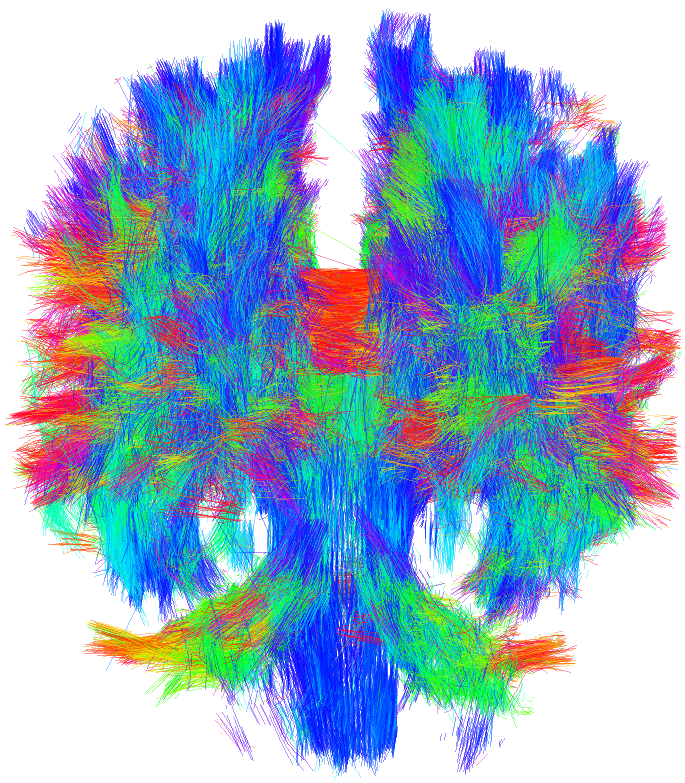}
     \hspace{.5cm}
     \includegraphics[width=0.34\textwidth]{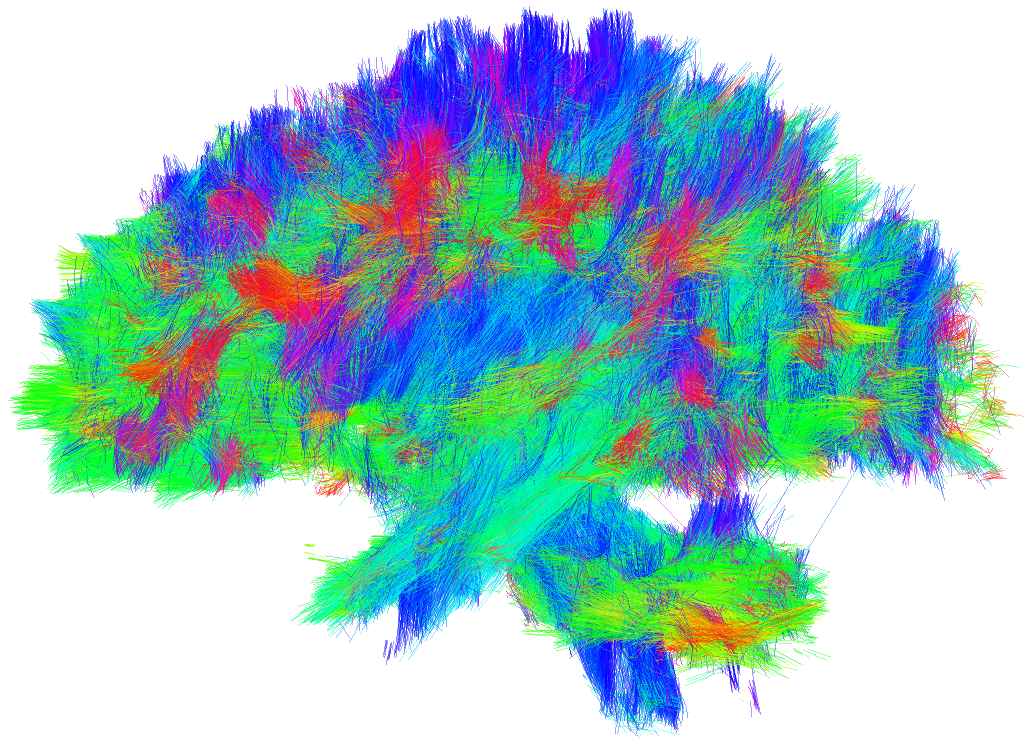}
     \hspace{.4cm}
     \includegraphics[width=0.16\textwidth]{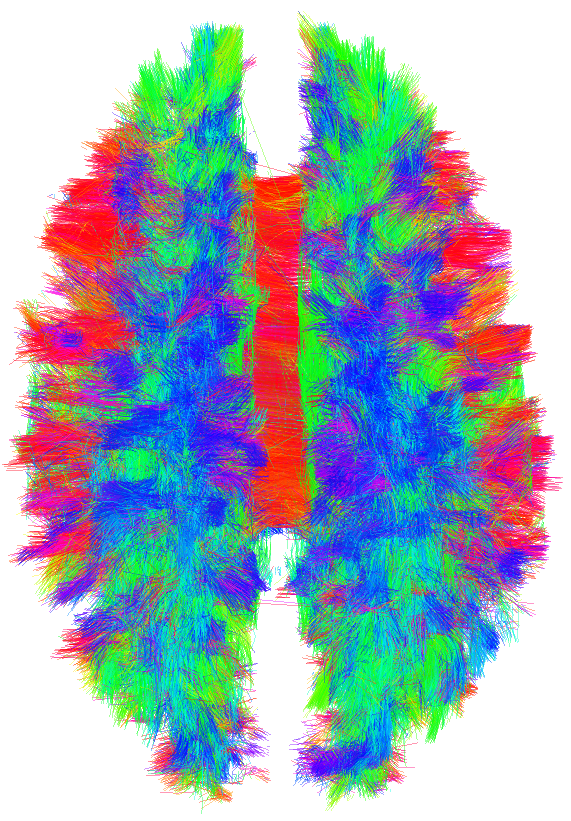}
     \vspace{.1cm}
     \includegraphics[width=0.83\textwidth]{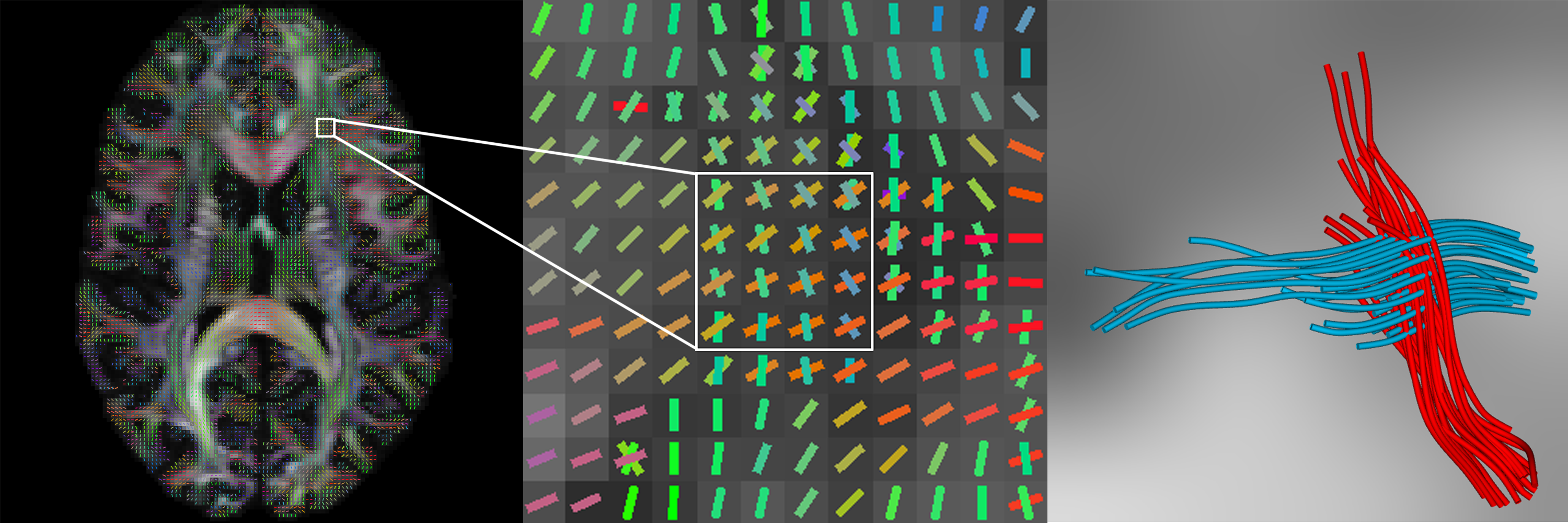}
        \caption{Top row, left to right: coronal, sagittal, and transversal view of the whole-brain connectome generated by the proposed method. Bottom left: axial view of the same subject. Bottom central: zoom-in of a $4\times4\times4$ crossing-fiber region. Bottom right: geodesic tractography by proposed method within the same window, the orientation of which matches the vector field in the bottom central panel.}
        \label{fig:3dbrain}
%   \vspace{-30pt}
\end{figure}

In this experiment, we validated our method's ability to estimate 3D crossing-fiber regions in brain DWI from several HCP subjects. We first reconstructed the vector fields through the GQI method~\cite{gqi} in DSI Studio with a diffusion sampling length ratio of 1.25. A whole-brain Riemannian metric was estimated by a CEDNN featuring 40, 30, and 40 dense layers in each dense block. The model was trained with an initial learning rate of $3\times10^{-4}$ for a total of $1\times10^4$ iterations. The top row in Fig.~\ref{fig:3dbrain} shows the resulting whole-brain connectome visualized using 3D Slicer via the SlicerDMRI plug-in~\cite{slicerdmri1}. There were 138,732 seed points cast in the white matter region for the generation of the geodesics. The color indicates the orientation of the fiber tracts: red (left/right), green (anterior/posterior), blue (superior/inferior). In the bottom row of Fig.~\ref{fig:3dbrain}, we showcase the ability of geodesic tractography with our estimated metric to successfully distinguish two crossing fibers: the forceps minor and frontal projection tracts. We stress that the previous approaches to geodesic tractography do not handle multiple fiber directions in a voxel, and thus cannot correctly handle crossing-fiber regions such as these. 

\section{Conclusion, Limitations and Future Work}\label{conclusion}
In this paper, we have shown for the first time how to leverage the flexibility of deep learning to model the shape of the human connectome by estimating a Riemannian metric of the brain manifold that faithfully represents the white matter connectivity. We
show that our proposed method outperforms any of the previously proposed methods in geodesic tractography by a large margin. In addition our approach solves the long-standing issue of these previous methods: the inability to recover crossing fibers with high fidelity. One limitation of the proposed method is that the generalization ability of the trained model to the unseen data is relatively weak. Nevertheless, we would point out that currently the largest human connectome datasets are in the order of only 1000s of subjects and thus we do not believe that the adaptivity of the models limits the applicability of our method, as the total training time for a single 3D brain is less than 30 minutes in our setup. With the ability to robustly and efficiently model the white matter of the brain as a Riemannian manifold, one can directly apply geometrical statistical techniques such as statistical atlas construction~\cite{campbell2021}, principal geodesic analysis~\cite{fletcher2004principal}, and longitudinal regression to precisely study the variability and differences in white matter architecture. 

% \newpage
\bibliographystyle{splncs04}
\bibliography{bibliography}
\end{document}